\documentclass[sigconf,screen,authoryear]{acmart}
\setcopyright{none}

\acmConference[WWW Companion '26]
{Companion Proceedings of the ACM Web Conference 2026}
{June 29--July 3, 2026}
{Dubai, United Arab Emirates}

\acmBooktitle{Companion Proceedings of the ACM Web Conference 2026 (WWW Companion '26), June 29--July 3, 2026, Dubai, United Arab Emirates}

\renewcommand\footnotetextcopyrightpermission[1]{}
\AtBeginDocument{%
  }
\usepackage[most]{tcolorbox}
\usepackage{graphicx}
\usepackage{float}        
\usepackage{listings}
\usepackage[utf8]{inputenc}
\usepackage[T1]{fontenc}
\usepackage{subcaption}
\usepackage{tikz}
\usetikzlibrary{shapes.geometric, positioning, fit, backgrounds, shadows, calc}
\usepackage{fontawesome5}

\lstdefinestyle{cleanR}{
basicstyle=\ttfamily\small,
columns=flexible,
keepspaces=true,
lineskip=2pt,
aboveskip=6pt,
belowskip=6pt,
breaklines=true,
showstringspaces=false
}

\usepackage{caption}
\usepackage{tcolorbox}
\tcbuselibrary{listings, skins}
\usepackage{xcolor}

\begin{document}

\title[Automating Computational Reproducibility in Social Science]{Automating Computational Reproducibility in Social Science: Comparing Prompt-Based and Agent-Based Approaches}

\titlenote{This is an updated version of the paper accepted for the \textit{Companion Proceedings of the ACM Web Conference 2026} (WWW Companion '26), DOI: \url{https://doi.org/10.1145/3774905.3795485}.}

\author{Syed Mehtab Hussain Shah}
\affiliation{
  \institution{GESIS - Leibniz-Institut für Sozialwissenschaften}
  \city{Cologne}
  \country{Germany}}
\email{mehtab.shah@gesis.org}

\author{Frank Hopfgartner}
\affiliation{
  \institution{University of Koblenz}
  \city{Koblenz}
  \country{Germany}}
\email{hopfgartner@uni-koblenz.de}

\author{Arnim Bleier}
\affiliation{
  \institution{GESIS - Leibniz-Institut für Sozialwissenschaften}
  \city{Cologne}
  \country{Germany}}
\email{arnim.bleier@gesis.org}

\begin{abstract}
Reproducing computational research is often treated as a straightforward matter of re-running shared code on shared data. In practice, published analyses frequently fail even when materials are available, due to missing dependencies, brittle paths, version mismatches, and—more challenging—gaps in computational logic. We investigate whether large language models (LLMs) and autonomous AI agents can automate the routine but labor-intensive work of repairing such failures, thereby lowering the practical barrier to verifying computational reproducibility.

We evaluate two automated repair paradigms in a controlled setting derived from five fully reproducible R-based social science studies. Starting from a ground-truth dataset, we inject realistic errors spanning execution-level issues, contextual code fixes, and structural logic omissions, and package them into 130 synthetic test cases. These test cases were systematically categorized based on the complexity and diversity of the injected errors. We then run (i) a prompt-based workflow that iteratively queries LLMs with structured prompts at increasing levels of contextual richness, and (ii) an agent-based workflow in which coding agents inspect project files, apply targeted edits, and re-run analyses until completion or a time limit. Repairs are only counted as successful if the execution outputs match the outputs of ground-truth dataset, not merely if the code executes.

Across prompt-based runs, reproduction success varies substantially (31–79\%), depending on failure complexity and the amount of context provided; richer context yields the largest gains on structurally complex cases. Agent-based workflows consistently achieve higher success (69–96\%) across all categories, suggesting that environment-aware, iterative agentic tool use provides a decisive advantage over solely prompt-based approaches. By isolating post-publication repair under controlled, systematic failure modes, our benchmark enables direct comparisons between prompt-based and agentic workflows and provides evidence that agentic repair systems can meaningfully reduce manual labour while improving the reliability of computational reproducibility in realistic social science research settings.
\end{abstract}



\maketitle
\section{Introduction}

Reproducing computational research still feels harder than it should be. Ideally, every published analysis would provide all code and data necessary to regenerate its results seamlessly and without requiring any difficult debugging or modifications~\citep{Claerbout1992ElectronicDG,Barba}. However, in practice this is often not the case. Empirical reviews suggest that only a small fraction of publications provide fully computationally reproducible analyses~\citep{chan2024,schoch2024}. Prior work has shown how common these problems are. For example, \citet{Hardwickerepro} manually examined 250 psychology articles and found that only a tiny minority of authors made analysis scripts available. Similarly, \citet{Trisovic2022} found that about 74\% of shared R scripts in replication datasets failed to run successfully in a clean environment. Large-scale surveys confirm the issue across disciplines; more than 70\% of researchers report struggling to reproduce published findings and over half cannot even reproduce their own work later on~\citep{Baker_2016}. Similar results have appeared in ecology and evolutionary biology, where only 15\% of meta-analyses shared usable data and code and most results still could not be reproduced~\citep{kambouris2024}. These patterns suggest a wider reproducibility problem that affects both verification and the long-term usefulness of scientific work.

\begin{figure*}[t]
  \centering
  \includegraphics[width=\textwidth]{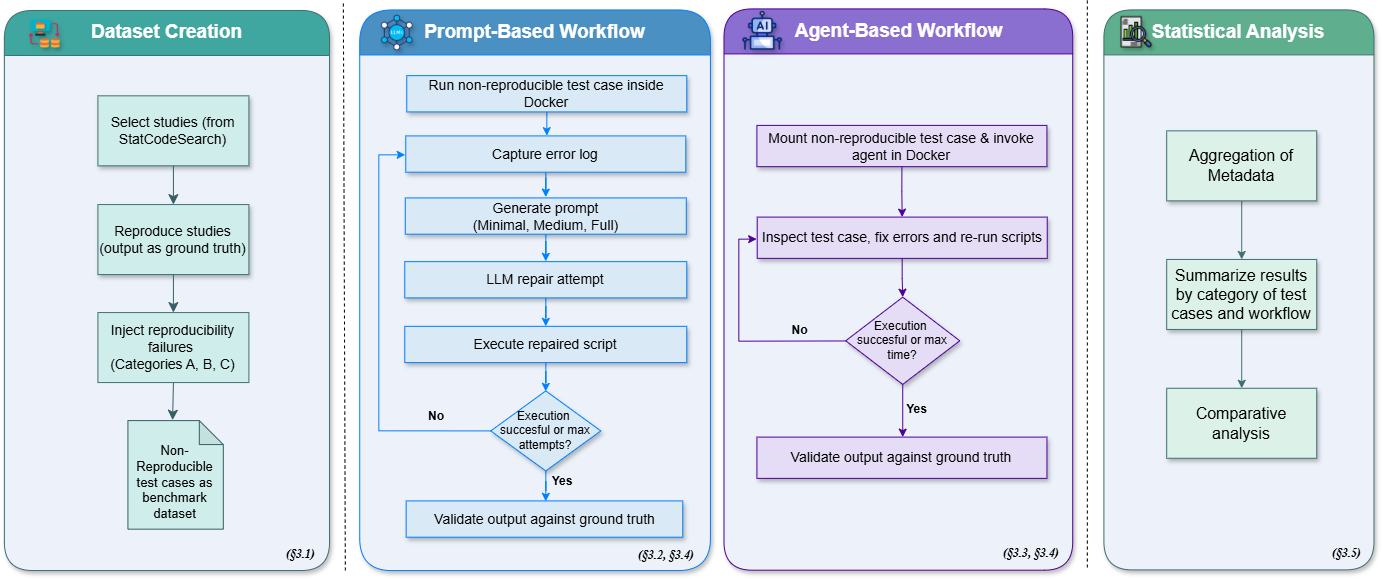}
  \caption{\textbf{Overview of the experimental framework.} First, we construct a synthetic benchmark dataset by injecting systematic failures into reproducible R studies to create a controlled testbed. Then, we evaluate two repair strategies: an iterative, prompt-based workflow and an autonomous agent-based workflow, both operating in isolated Docker environments. Finally,  results were aggregated by category of test cases and workflow to enable comparative statistical analyses. Section markers in parentheses (e.g., §3.1) indicate where each stage is described in detail.}
  \label{fig:workflow}
\end{figure*}

At the same time, AI is transforming how researchers write and execute computational analyses. Modern large language models (LLMs) can read scripts, suggest revisions, and explain errors. Some studies argue that they may help reduce the burden of debugging and might even handle some reproducibility tasks independently~\citep{Zhang_2025,Reason_2024}. The idea is that instead of requiring human experts to manually go through every analysis step and debug problems, we could prompt LLMs to carry out these steps automatically or at least to assist researchers more effectively. However, it is not yet clear how far these tools can go. LLMs often make plausible but incorrect suggestions, miss analytical logic, or ignore domain-specific assumptions that human experts would typically notice~\citep{WYSOCKA2024104724,llmhallucination}. Systematic empirical evidence in the social sciences remains limited regarding whether LLMs and AI agents can reliably reproduce published results, and under which conditions their success or failure occurs.

We address this gap by introducing a benchmark dataset and conducting two separate reproducibility experiments. One experimental workflow uses a prompt-based approach with varying levels of context. The other employs an agent-based approach that operates directly on the code. Both workflows are evaluated in isolated environments and use the same dataset with identical test cases. The goal is to measure how well each approach repairs non-executable research code and if these tools can recover the original published results. The aim is to gain empirically grounded insights into where AI tools fit within computational reproducibility studies in the social sciences and what their current limitations are. Building on this motivation, we structure this report around the following research questions: \\
\textbf{RQ1:} To what extent can LLM–based workflows automatically repair published computational social science analyses? \\
\textbf{RQ2:} How does the performance of prompt-based LLM repair workflows compare to agent-based workflows? \\
\textbf{RQ3:} How does the complexity of reproducibility failures affect the success of automated repair approaches? \\
\textbf{RQ4:} How does the amount and type of contextual information influence the ability to repair non-executable analyses? \\

We report on several contributions. First, we present a novel synthetic benchmark dataset designed to assess computational reproducibility in R-based analyses typical of computational social science. This benchmark consists of fully reproducible studies paired with systematically injected, realistic errors that reflect common barriers encountered when reproducing published research. Second, we propose a prompt-based reproducibility repair workflow that uses different prompts for LLMs to diagnose and repair non-executable analysis scripts under varying levels of contextual information. Third, we introduce an agent-based reproducibility repair workflow in which standard off-the-shelf coding agents are used to inspect files containing computational analyses, modify code, iteratively re-run analyses, and reason about errors in a more interactive manner. Finally, we report on an empirical evaluation of both workflows on the synthetic benchmark, comparing their effectiveness in restoring computational reproducibility across different classes of errors and validating whether successful repairs restore the original analytical results. This work establishes a controlled benchmark and evaluation framework for studying AI-assisted reproducibility repair, enabling systematic comparison between prompt-based and agent-based approaches. 

\section{Related Work}

Computational reproducibility is widely regarded as a necessary requirement for credible and cumulative science \cite{national2019}; however, empirical evidence consistently shows that it remains difficult to achieve in practice. Large-scale surveys and audits demonstrate that shared code and data are frequently missing, incomplete, or non-executable, even in fields with strong computational traditions \cite{Baker_2016,Hardwickerepro,schoch2024,chan2024}. Similar findings appear outside social science: attempts to computationally reproduce published meta-analytic results often fail despite nominal artifact availability, underscoring that transparency alone is insufficient \cite{kambouris2024}. Collectively, this work highlights a significant gap between stated reproducibility norms and empirical reality.

To address these challenges, research has focused on stabilizing computational environments and preserving execution context. Containerization and environment encapsulation have become standard approaches for mitigating dependency and configuration drift, particularly in R-based workflows \cite{boettiger2017introductionrockerdockercontainers}. Complementary tools emphasize provenance capture and packaging, bundling scripts, data, and dependencies into self-contained artifacts that can be re-executed elsewhere \cite{ReproZip}.

Recently, LLMs have emerged as potential tools for reproducing non-reproducible computational workflows after publication. Benchmarks in code generation and program understanding show that modern LLMs can synthesize functional code from natural language specifications \cite{chen2021evaluatinglargelanguagemodels,diera2023gencodesearchnetbenchmarktestsuite}. Dedicated studies of debugging and code repair demonstrate that LLMs can fix non-trivial defects, although performance remains imperfect and sensitive to task structure \cite{tang2024coderepairllmsgives,Saboor_Yaraghi_2025}. Agent-based approaches further improve repair capability by structuring interaction around execution feedback and iterative reasoning \cite{yang2025coastenhancingcodedebugging}.

Several recent benchmarks evaluate LLMs and agents directly on scientific reproducibility and replication tasks. CORE-Bench frames computational reproduction as an interactive agent task requiring execution, inspection, and output validation across difficulty levels \cite{siegel2024corebenchfosteringcredibilitypublished}. PaperBench evaluates end-to-end reproduction of machine learning research from paper descriptions, revealing sharp performance drops as executable structure is removed \cite{starace2025paperbenchevaluatingaisability}. In the social sciences, REPRO-Bench evaluates reproducibility by asking agents to inspect a paper together with its associated code and data, and determine whether the reported results can be reproduced \cite{hu2025reprobenchagenticaisystems}. Similarly, \citet{zhang2026paperreproautomatedcomputationalreproducibility} present PaperRepro, a two-stage multi-agent reproducibility assessment pipeline that separates execution from evaluation to better capture results and improve over prior baselines. Together, these works highlight the potential of AI for scientific reproducibility, but they frame it as an assessment task, not a repair task: the goal is to determine whether results can be reproduced given existing artifacts. They therefore do not focus on the complementary challenge of repairing analysis workflows that fail to execute or reproduce results. In particular, existing studies do not explicitly address post-publication reproducibility failures, where code and data are available but the analysis still does not work without extra manual fixing.

This leaves open the question of how such failures can be systematically studied and compared across different automated repair approaches. To the best of our knowledge, no prior work has introduced a synthetic benchmark dataset specifically designed to assess computational reproducibility in R-based social science studies. Moreover, existing studies do not provide a systematic comparison between prompt-based and agent-based repair workflows in this context, nor do they examine the impact of varying levels of prompt context on reproducibility outcomes.
Taken together, these gaps motivate a focused comparison of prompt-based and agent-based approaches for repairing non-executable computational social science studies.

\section{Methodology}
We conduct controlled experiments to evaluate whether LLMs, specifically prompt-based and agent-based approaches, can restore non-ex\-ecutable research code and aid in achieving computational reproducibility. To this end, we constructed a controlled study of reproducibility. Our methodology operationalizes the research questions posed in Section 1 through a new synthetic benchmark, as well as by systematically varying the models and harnesses used, error complexity, and contextual information to assess the ability of LLMs to repair non-executable research code under controlled conditions. We begin by describing the synthetic benchmark; next, we introduce the prompt-based and agent-based workflows, followed by a description of the validation steps taken to ensure that the AI-driven repair attempts were indeed successful. Figure~\ref{fig:workflow} illustrates our methodological design.

\subsection{Synthetic Benchmark Creation}
\label{sec:dataset}

We created our synthetic benchmark dataset in a structured process. 
First, we curated a set of fully computationally reproducible publications as a ground-truth dataset. 
Then, we created synthetic test cases by producing modified copies of the ground-truth publications, where each copy contains one or more deliberately introduced errors. These modified versions represent non-reproducible variants of the original, fully reproducible publications.

To curate the set of fully computationally reproducible publications, we defined the following three selection criteria for candidate papers: 1) a paper must be openly accessible, 2) all necessary code and data appendices must be equally accessible, 3) and executing the code together with the provided data must reproduce the original results reported in the corresponding publication. To verify reproducibility, we followed the workflow described by \citet{saju_holtdirk_mangroliya_bleier_2025}. The computational aspects of reproducibility testing were done in Rocker \citep{boettiger2017introductionrockerdockercontainers} (rocker/r-ver:4.4.1) Docker environments. Each container was allocated 8~GB of RAM and 8 CPU cores. With these criteria in place, we evaluated the subset of StatCodeSearch \citep{diera2023gencodesearchnetbenchmarktestsuite} described in \citet{saju_holtdirk_mangroliya_bleier_2025} and identified five papers that satisfy all criteria.

\begin{figure}[t]
  \centering

  \begin{subfigure}[b]{0.45\columnwidth}
    \centering
    \resizebox{\linewidth}{!}{%
      \begin{tikzpicture}[
        font=\sffamily,
        labeltext/.style={font=\scriptsize, align=center, text width=2.5cm} 
      ]

        \draw[->, thick] (0,0) -- (4.7,0)
          node[midway, below=0.35cm, font=\fontsize{10}{9}\selectfont]
          {Diversity};
        
        \draw[->, thick] (0,0) -- (0,6.5)
          node[midway, rotate=90, above=0.35cm, font=\fontsize{9}{9}\selectfont] 
          {Difficulty};

        \filldraw[fill=cyan!40] (1.2,1.6) circle (0.25cm);
        \node[align=center, below, font=\small] at (1.2,1.4)
          {Category A\\(Execution Errors)};

        \filldraw[fill=blue!40] (2.2,3.5) circle (0.25cm);
        \node[align=center, below right, font=\small] at (1.0,3.3)
          {Category B\\(Contextual Errors)};

        \filldraw[fill=purple!40] (3.5,5.6) circle (0.25cm);
        \node[align=center, below, font=\small] at (3.5,5.4)
          {Category C\\(Structural Errors)};

        \draw[dashed, gray!60] (1.2,1.6) -- (1.2,0);
        \draw[dashed, gray!60] (1.2,1.6) -- (0,1.6);

        \draw[dashed, gray!60] (2.2,3.5) -- (2.2,0);
        \draw[dashed, gray!60] (2.2,3.5) -- (0,3.5);

        \draw[dashed, gray!60] (3.5,5.6) -- (3.5,0);
        \draw[dashed, gray!60] (3.5,5.6) -- (0,5.6);

      \end{tikzpicture}%
    }
    \caption{Categories of errors} \label{fig:error_cat}
  \end{subfigure}%
  \hspace{0.15cm}
    \begin{tikzpicture}[baseline=0pt]
    \draw[densely dotted] (0,0) -- (0,6.2cm); \end{tikzpicture}
  \hspace{0.15cm}
  \begin{subfigure}[b]{0.45\columnwidth}
    \centering
    \resizebox{\linewidth}{!}{%
      \begin{tikzpicture}[
          node distance=0.25cm and 0.15cm,
          font=\sffamily,      
          /utils/exec={\usetikzlibrary{shapes.geometric, fit, backgrounds, shadows, positioning}},
          file/.style={
              rectangle, draw=gray!40, top color=white,
              bottom color=gray!5, drop shadow,
              minimum width=1.0cm, minimum height=1.4cm,
              rounded corners=2pt, align=center
          },
          database/.style={
              cylinder, shape border rotate=90, aspect=0.25,
              draw=orange!60!black, top color=white, bottom color=orange!10,
              drop shadow, minimum width=1.0cm, minimum height=1.4cm,
              align=center
          },
          folder/.style={
              draw=blue!50!black, fill=blue!5, dashed,
              line width=0.9pt, rounded corners=4pt, inner sep=0.2cm
          }
      ]
        \node[file] (paper) {
            \\[-0.2mm]
            \textbf{\small P} \\[-0.8mm]
            \textbf{\tiny paper.md} \\[-0.8mm]
            \fontsize{4}{4}\selectfont (Context)
        };
        \node[database, right=0.3cm of paper,yshift=-0.1cm] (data) {
            \textbf{\small D} \\[-1mm]
            \textbf{\tiny data.csv} \\[-1mm]
            \fontsize{4}{4}\selectfont (Input)
        };
        \node[file, below=0.3cm of paper, draw=red!50, bottom color=red!5, line width=0.9pt] (code) {
            \textbf{\small </>} \\[-0.3mm]
            \textbf{\tiny script.R} \\[-1mm]
            \fontsize{4}{4}\selectfont (Injected \\[-1mm]  \fontsize{4}{4}\selectfont Errors)
        };
        \node[file, right=0.3cm of code] (utils) {
            \textbf{\small \{ \}} \\[-0.3mm]
            \textbf{\tiny utils.R} \\[-1mm]
            \fontsize{4}{4}\selectfont (Supporting \\[-1mm] \fontsize{4}{4}\selectfont Scripts)
        };
        \begin{scope}[on background layer]
          \node[folder, fit=(paper)(data)(code)(utils),
                label={[anchor=south west, text=blue!50!black, font=\bfseries\tiny, ]}] (container) {};
        \end{scope}
      \end{tikzpicture}%
    }
    \vspace*{0.2cm}
    \caption{Test case structure} \label{fig:sample_structure}
  \end{subfigure}

 \caption{\textbf{Overview of the synthetic reproducibility benchmark.} \textbf{(a)} Classification of error types plotted against error diversity and repair difficulty, ranging from simple execution errors (Category A) to complex structural logic gaps (Category C). \textbf{(b)} The structural organization of a single synthetic test case, containing the original publication as context, input data, support scripts, and the analysis script with multiple injected errors.}
  \label{fig:side_by_side}
\end{figure}
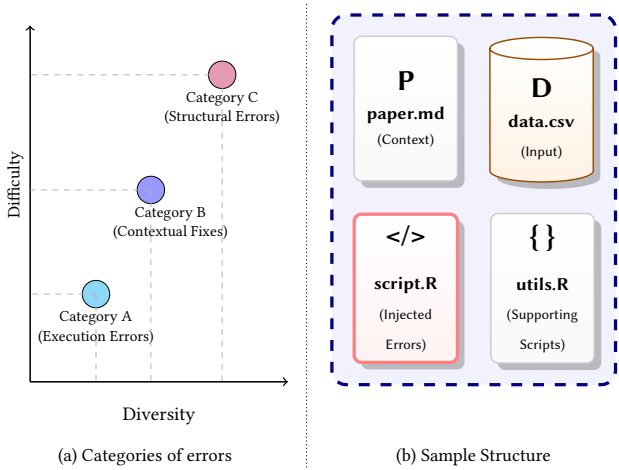

To generate the synthetic test cases, we started by creating and injecting errors into copies of the code appendices of our ground-truth dataset. To better mimic real-world conditions, we generated our synthetic errors based on findings from large-scale studies identifying the most common reasons why computational reproducibility fails \citep{saju_holtdirk_mangroliya_bleier_2025, Trisovic2022, costa2025datasetcomputationalreproducibility}. They include problems such as missing packages, incorrect file paths, missing objects or functions, shared library loading issues, package installation failures, and file-reading errors. Furthermore, in real-world settings, researchers are typically faced with combinations of errors that prevent computational reproducibility \citep{Liu2019SuccessesAS, breznau2025reliability}. To reflect this in our benchmark, we constructed synthetic test cases by combining multiple injected errors, including combinations spanning different categories. Each test case contains either a single synthetic injected error or a set of multiple injected errors. To structure the benchmark, we organized the test cases into three categories that capture two main factors: first, the diversity of potential issues encountered during a reproduction attempt; and second, the difficulty of achieving a successful re-run of the computations. The three categories of test cases are as follows:

\begin{itemize}
\item \textbf{Category A.} Test cases containing errors that affect execution without altering the analytical structure of the code. Examples include wrong file paths, missing packages or simple typos in the code. This category encompasses errors that are potentially easy to fix but require some knowledge of the required packages as well as insight into the file system of the execution environment.

\item \textbf{Category B.} Test cases that include errors requiring more nuanced, code-level fixes and reasoning about the analysis. Examples include incorrect or outdated packages, several syntax errors, missing variables or small missing parts of the code. In addition to these errors, these test cases also incorporate combinations of one or more errors from Category A test cases, increasing the overall complexity of Category B.

\item \textbf{Category C.} Test cases that involve structural issues, where the analytical logic is incomplete or missing. These include scenarios with absent functions, incomplete code blocks, or combinations of multiple error types - such as path issues together with missing variables or syntax errors that affect multiple files simultaneously. Category C test cases also contain combinations of simpler and intermediate errors used in Categories A and B test cases, capturing the diversity and complexity of issues that require deep understanding of the intended computations to reproduce successfully.
\end{itemize}
Figure~\ref{fig:error_cat} provides a graphical overview of the three categories of synthetic test cases.

 Across all five papers, this resulted in a total of 130 unique synthetic test cases in our benchmark. Figure~\ref{fig:sample_structure} illustrates the structure of an individual synthetic test case consisting of the non-executable code with multiple synthetic errors, accompanying utility functions, required data, as well as a Markdown representation of the publication the code belongs to. The diversity of error types embedded within these test cases enables a controlled evaluation of how error complexity, as defined by varying levels and combinations of errors, influences the success of automated repair methods (RQ3).

\subsection{Prompt-based Repair Workflow}
\label{sec:prompt_based}

In the first experiment, we evaluated a prompt-based repair pipeline in which large language models were used with structured prompts of varying contextual detail to iteratively detect and correct errors in test cases of our benchmark. Docker containers were used to execute the code and record the logs for the experiments. We used the Docker environment and system configuration described in Section~\ref{sec:dataset} ensuring consistency and comparability across experiments.
When execution failed, the failing script and associated error logs were collected and used to construct a prompt for the selected model. 
For each execution-repair loop - one cycle in which the model proposes updates to the script and the script is re-run to check for errors — a new, isolated container was instantiated to ensure a consistent runtime environment. Stochasticity therefore arises only from model inference and is reflected in aggregate success rates. Every run produced a standardized set of artifacts, including raw console logs, structured metadata summarizing the run, and the final state of the repaired working directory.

We evaluated three models, GPT-4o (release 2024-08-06) \cite{openai_structured_outputs_2024}, Gemini~2.5~Pro (release preview-06-05) \cite{gemini_blog}, and Qwen3-Coder-480B‑A35B \cite{qwen3coder} using the same repair loop to ensure comparability. Furthermore, prompts were generated at three levels of contextual richness. Based on early trials on a small set of tasks, we gradually adjusted the prompt templates to ensure that the repair process worked reliably from start to finish and that the models clearly understood what they were asked to do. This resulted in three well-defined prompt versions that shared the same core instructions but differed in how much supporting information they included.

\begin{itemize}
\item \textbf{Minimal prompt.} Includes the script, the log-file from the previous run, and a request to fix any existing issues.
\item \textbf{Medium prompt.} Includes the minimal prompt plus the text of the publication in Markdown, giving the model access to the analysis as described in the respective papers.
\item \textbf{Full prompt.} Includes the medium prompt plus all supporting scripts for the respective paper and detailed instructions on how to perform the repair task.
\end{itemize}
The complete templates for the Minimal, Medium, and Full prompts are provided in Appendix~\ref{app:prompts}.

For each attempt, the model was requested to return an updated version of the script, which was then re-executed in the same container. If execution failed again, the new error output was appended to the prompt and the process repeated, up to a maximum of five iterations. A run was considered successful only if the script completed without errors and reproduced the original ground-truth outputs. A detailed discussion of this validation is provided in Section~\ref{sec:validation}. Across all 130 test cases described in Section~\ref{sec:dataset}, all three models and three prompt types were evaluated, resulting in 1,170 unique repair runs. By varying prompt context while holding execution conditions constant, this workflow allows us to examine both the overall effectiveness of prompt-based repair (RQ1) and the role of contextual information in addressing analytically complex errors (RQ4). 

\subsection{Agent-based Repair Workflow}
The second experiment evaluated an agentic repair approach in which AI agents operated directly inside the containerized environment. This workflow reused the same set of synthetic test cases; however, instead of our custom prompts, we employed off-the-shelf coding agents with access to the execution environment. These agents could inspect files, modify code, and iteratively re-run analyses. To evaluate the agent-based repair workflow, we focused on a single model, Qwen3-Coder. We selected Qwen3-Coder-480B‑A35B due to its strong performance on real-world coding benchmarks\footnote{See, e.g., SWE-bench, which evaluates a model’s ability to resolve practical coding tasks: https://swebench.com (accessed on 15-08-2025)}.

We embedded the OpenCode\footnote{https://opencode.ai} and Claude Code\footnote{https://code.claude.com} agents directly into the Docker image. Both agents were configured to use Qwen3-Coder-480B-A35B as their underlying language model via OpenRouter\footnote{https://openrouter.ai}. Since Claude Code natively supports only Anthropic models, we employed Claude Code Router\footnote{A routing layer that enables Claude Code to interface with third-party model providers: https://musistudio.github.io/claude-code-router/} to enable compatibility with the selected model. By fixing the underlying language model, this setup ensures that observed differences in performance — between prompt-based and agent-based approaches — can be attributed to differences in workflow and control strategy, such as how they manage iterations and tool use, instead of model choice. For each test case, we launched a fresh container with the corresponding test case directory mounted as the working directory and executed the selected agent in a headless, non-interactive mode using the prompt shown in Appendix~\ref{app:prompt}. The ground-truth outputs were mounted separately in the container to enable a final reproducibility check\footnote{We visually inspected the agents’ logs to verify their behavior; in all runs, the agents did not read the ground-truth data before completing the repair process.}. The agents were free to explore the workspace, understand the paper, edit files, and re-run the R scripts. The agents were run until the reproduction attempt succeeded or a maximum time limit of 20 minutes was reached. All agent runs again used the Docker environment and system configuration described in Section~\ref{sec:dataset}. This agent-based workflow is designed to evaluate whether direct interaction with the execution environment provides practical advantages over prompt-based LLM workflows when repairing complex reproducibility failures (RQ2).

\subsection{Validation}
\label{sec:validation}
After each repair attempt, whether conducted via the prompt-based or an agent-based workflow, the output of the repaired script was automatically compared against the ground-truth dataset of the original study and the Markdown version of the publication. Based on the comparison results, a classification label Reproduced, or Not Reproduced was assigned to that run. Here, \textbf{Reproduced} means the script ran to completion and produced outputs that matched the stored ground-truth. Whereas \textbf{Not Reproduced} means the script either failed to run or outputs differed from the ground-truth. This validation is important because, as noted by \citet{WYSOCKA2024104724}, a model may appear to solve a problem and write code that runs without errors, yet still produce analytically incorrect outputs.

\subsection{Analysis}

The prompt-based as well as the agent-based workflows produced structured records for each repair attempt, including the test case identifier, error type, test case category, model or agent used, number of attempts, execution time, and final reproduction outcome. 

These records were aggregated to enable a systematic comparison of the different approaches. The aggregated data forms the basis for the results presented in Section~\ref{sec:results}, where we evaluate the effectiveness and efficiency of the workflows across different test case categories and error types.

\section{Results}
\label{sec:results}
We evaluated performance using the same metrics across workflows, focusing on reproduction success rates broken down by test case category, prompt type, and model or agent used. To this end, we aggregated results by workflow, model or agent, test case category, and reproducibility status to assess whether successful repairs preserved the original analytical intent. This section presents the empirical findings of our evaluation across prompt-based and agent-based workflows. The analysis is structured around our research questions: differences between workflows (RQ1–RQ2), sensitivity to error complexity (RQ3), and the influence of contextual information (RQ4). We first report results for the prompt-based workflow across different models and test cases. We then present the results for the agent-based workflow. Finally, we compare both approaches in terms of overall performance, robustness, and their response to varying task conditions.

\subsection{Results from the Prompt-based Workflow}
\begin{figure}[t]
  \centering
  \includegraphics[width=\columnwidth]{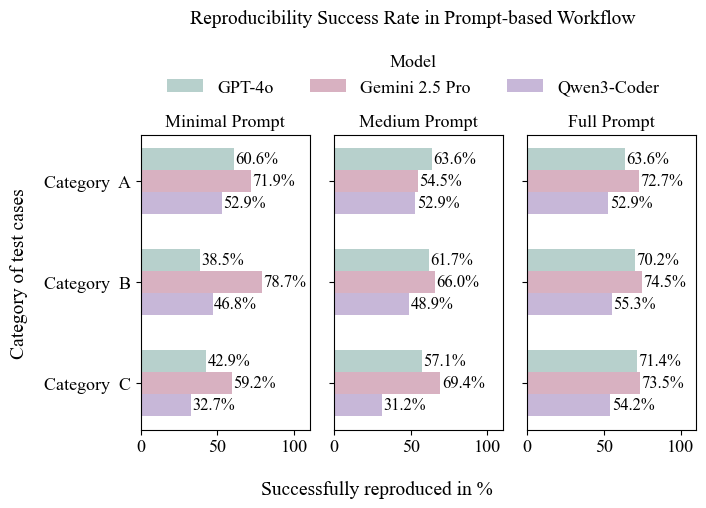}
  \caption{\textbf{Prompt-based LLM Repair Workflow}. Success rates (\%) of GPT-4o, Gemini 2.5 Pro, and Qwen3-Coder across test case categories (y-axis) and prompt types (x-axis). Each prompt level shows a grouped bar chart of model performance for that error–prompt combination, illustrating how error complexity and contextual information affect automated reproducibility.}
  \Description{Bar chart showing success rates of GPT-4o, Gemini 2.5 Pro, and Qwen3-Coder for various error complexities and prompt types.}
  \label{fig:Success_matrix_detailed}
\end{figure}

By examining success rates across test case categories and prompt types (Figure~\ref{fig:Success_matrix_detailed}), we see that each model handles reproducibility challenges in a distinct way, with some models performing better on certain types of errors or prompt conditions than others.
\\[1em]
\noindent\textbf{Gemini 2.5 Pro.}
Gemini 2.5 Pro consistently leads across nearly all conditions. With the minimal prompt, it already achieves solid performance on Category A (71.9\%) and Category B (78.7\%) test cases, and maintains relatively strong outcomes on Category C (59.2\%). Interestingly, performance is not strictly monotonic with prompt context: while the medium prompt improves some scores in Category C, other categories see a slight drop, consistent with recent evidence that additional, not required context can degrade LLM performance even when the provided context is accurate \citep{du2025contextlengthhurtsllm,zhang2025recursivelanguagemodels}. However, the full prompt generally restores and enhances performance best, particularly for Category C test cases where errors require restoration of analytical logic, with success rates reaching 73.5\%.
\\[1em]
\noindent\textbf{GPT-4o.}
The results for GPT-4o show that success is highly dependent on prompt richness. Under a minimal prompt, the model struggles with Category C test cases, achieving only a 42.9\% success rate. As context increases, however, performance improves significantly. The medium prompt boosts success rates across all error complexities, and the full prompt pushes the success rate for Category C test cases up to 71.4\%, nearly matching Gemini 2.5 Pro. This pattern illustrates that GPT-4o is a strong model, but one that requires sufficient context in the prompt to reach its potential. Without that context, GPT-4o tends to underperform, especially on tasks that demand analytical problem solving and logic building.
\\[1em]
\noindent\textbf{Qwen3-Coder.}
Qwen3-Coder presents a profile consistent with more lightweight open models. Category A test cases are handled reliably (52.9\%), even with the minimal prompt, suggesting a solid baseline for straightforward fixes. However, performance drops sharply as complexity rises: success rates for Category B test cases range from 46.8\% to 55.3\%, while Category C test cases range from 32.7\% to 54.2\%. Qwen3-Coder also appears to improve with more context, especially for Category C test cases.
\\

These patterns highlight that GPT-4o has strong code repair capabilities when provided with sufficient context, while Gemini 2.5 Pro demonstrates robust performance even with sparse minimal prompts. Qwen3-Coder offers some utility on simpler errors but has clear limitations on complex cases. The findings also reveal that LLMs are capable of meaningfully repairing erroneous research code and restoring functionality in a substantial portion of real-world cases, but their effectiveness depends heavily on the nature of the error, the structure and clarity of the original research workflow, as well as the amount and type of context that is provided to the model. These findings are consistent with those reported by \citet{sun2025empiricalevaluationlargelanguage}, who similarly observe that LLM performance in coding is highly sensitive to task complexity and prompt/context quality. Overall, these findings provide initial answers to RQ1 and RQ4, showing that while prompt-based workflows can restore reproducibility in a substantial share of cases, their effectiveness depends strongly on both error complexity and the availability of relevant contextual information.

\subsection{Results from the Agent-based Workflow}

\begin{figure}[t]
  \centering
  \includegraphics[width=\columnwidth]{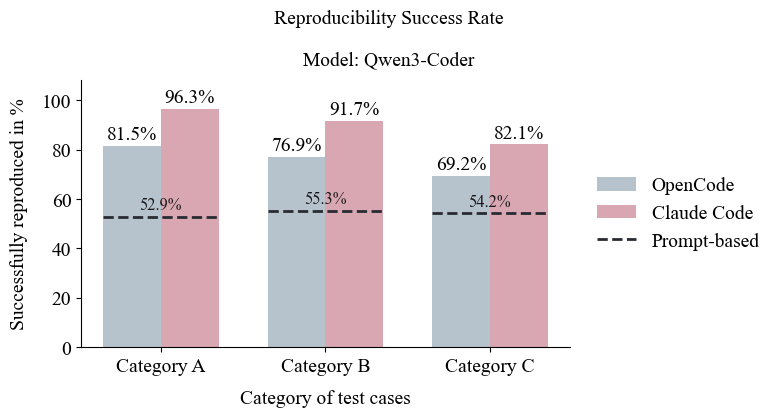}
  \caption{\textbf{Qwen3-Coder Performance in Agent-based and Prompt-based Workflow.} Percentage of successfully reproduced cases using Qwen3-Coder across categories. Bars show agent-based workflow performance with OpenCode and Claude Code, while dashed horizontal lines indicate for comparison the success rates of the prompt-based workflow using Qwen3-Coder with the full-context prompt.}
  \label{fig:agentllm}
\end{figure}

The following section presents the outcomes of the agent-based workflow for restoring computational reproducibility. We evaluate how each agent performed in restoring non-executable scripts and reproducing the original study results across the different test case categories.
\\[1em]
\noindent \textbf{OpenCode}. Figure~\ref{fig:agentllm} shows the results for the OpenCode coding agent using the Qwen3-Coder model. The figure illustrates the success rates across the test case categories (Category A, Category B, Category C).
The OpenCode agent was able to successfully repair a majority of scripts across all three test case categories.  For Category A test cases, 81.5\% of the scripts were fully repaired and produced results identical to the original study. Category B test cases, which included missing libraries or small sections of code, were successfully repaired in 76.9\% of cases. Even the complex errors (Category C), which involved reconstructing missing logic or resolving complex interactions across multiple parts of the code, were correctly fixed in 69.2\% of the scripts. The OpenCode agent demonstrated strong capabilities in systematically identifying and resolving errors while restoring the original analytical outcomes and making the studies reproducible.
\\[1em]
\noindent \textbf{Claude Code}. As shown in figure~\ref{fig:agentllm}, the Claude Code agent exhibited even higher repair performance compared to OpenCode. Errors in test cases of Category A were fully resolved in 96.3\% of scripts, restoring the intended results almost universally. For Category B test cases, 91.7\% of scripts were repaired successfully, while even the test cases containing the most complex errors (Category C) were correctly fixed in 82.1\% of cases. These results show that Claude Code was highly effective in handling more complex and logic-intensive errors, systematically identifying the causes of failure and generating corrected scripts that matched the ground-truth outputs.

\subsection{Prompt-based vs Agent-based Workflows}

To better understand how the prompt-based workflow performs in comparison to the agent-based workflow, we compared success rates using the same model i.e., Qwen3-Coder with OpenCode and Claude agents.
Figure~\ref{fig:Success_matrix_detailed} shows the success rate on Qwen3-Coder for different categories of test cases in the prompt-based workflow. Qwen3-Coder performed consistently best with full prompt type, therefore, we focus on comparing the full-prompt results with the agent-based workflow.
\\[1em]
\noindent \textbf{Prompt-based workflow}: In the prompt-based workflow, Qwen3-Coder showed moderate performance across error types. For Category A test cases, success rates were consistent at 52.9\%. Category B achieved 55.3\% with the full-context prompt, and Category C test cases, which were the most challenging, peaked at 54.2\% under the full-context prompt type, where the model had access to all scripts, the original paper, and detailed instructions (see Figure~\ref{fig:agentllm}).
\\[1em]
\noindent \textbf{Agent-based workflow}: When we used the same Qwen3-Coder model within agent-based workflows, success improved notably. Using the OpenCode agent, Category A test cases were reproduced 81.5\% of the time, Category B 76.9\% and Category C test cases 69.2\% (Figure~\ref{fig:agentllm}). Similarly, with the Claude agent, performance was even higher, with 96.3\% of success in Category A, Category B at 91.7\% and Category C test cases reaching 82.1\% (see Figure~\ref{fig:agentllm}).

\begin{figure}[t]
  \centering
  \includegraphics[width=0.9\columnwidth]{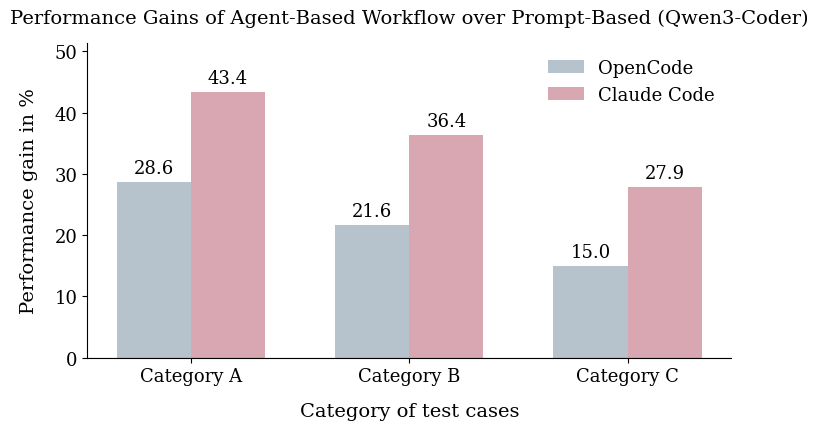}
  \caption{\textbf{Improvement of Agent-based Workflows over Prompt-based.} Absolute improvement in reproducibility success rates achieved by agent-based workflows relative to the prompt-based workflow using Qwen3-Coder in full-context prompt mode. Bars show the difference in percentage points, representing the absolute increase in success rates for each agent (OpenCode and Claude Code) compared to the prompt-based workflow.}
  \label{fig:gains}
\end{figure}

To highlight the specific contribution of the agent-based workflow, we calculated the net performance gain of each agent relative to the standalone prompt-based workflow using the same model (Qwen3-coder) with full-context prompts, as illustrated in Figure~\ref{fig:gains}. For the simplest errors (Category A), the OpenCode agent improved success rates by 28.6 percentage points, while the Claude Code agent delivered an even larger boost of 43.4 percentage points. For intermediate errors (Category B), gains were 21.6 percentage points with OpenCode and 36.4 percentage points with Claude. Even on the most complex structural errors (Category C), agent-based workflows continued to outperform the prompt-based workflow, with 15.0 percentage points gained using OpenCode and 27.9 percentage points with Claude.

These results demonstrate that the agent-based workflow consistently outperforms the isolated prompt-based API approach. The agent-based workflow allows models to handle more complex cases effectively, likely due to the agents’ ability to actively explore the workspace, analyze error logs, iteratively modify scripts, and re-run analyses until a correct solution is found. In contrast, prompt-based workflows are limited and cannot interact with or adapt to the workspace, which constrains their ability to resolve more complex situations that involve multiple errors at the same time. The comparison highlights that the observed performance gap between workflows is not solely model-dependent but is equally driven by differences in interaction and control afforded by agent-based execution, thereby directly addressing RQ2.

\section{Conclusion and Future Work}

In this paper, we reported on work in progress exploring the potential of large language models and agent-based approaches to support computational reproducibility in the social sciences. Using a controlled synthetic benchmark of R-based studies with systematically injected failures, we evaluated two automated repair workflows, a prompt-based approach and an agent-based approach, under validation conditions where the ground-truth is known. Our findings offer clear answers to the research questions posed at the outset of this study.

Regarding RQ1 (extent of automated repair) and RQ3 (impact of error complexity), our results suggest that prompt-based workflows can successfully repair and reproduce a non-trivial share of corrupted analyses for which reproduction of computational results would have been impossible without prior repair, particularly in cases involving execution-level and moderately complex errors. However, their effectiveness is strongly conditioned on the availability and structure of contextual information, and performance degrades substantially when errors require reconstructing missing analytical logic (Category C). Concerning RQ4 (influence of contextual information), we found that the effectiveness of prompt-based repair is strongly conditioned on the availability and structure of context. Providing the full text of the publication and supplementary scripts significantly improved success rates for complex test cases, whereas minimal context was often insufficient for logic-heavy repairs. This confirms that while LLMs can reduce manual debugging effort, they remain highly sensitive to prompt design and context selection. Finally, addressing RQ2 (comparison of workflows), our analysis shows that agent-based workflows consistently outperform prompt-based LLM approaches across all categories of test cases, with particularly strong gains for complex, multi-step failures. By allowing models to interact directly with the execution environment—inspecting files, iteratively modifying code, and re-running analyses—agentic systems appear better suited to handling the practical realities of computational reproducibility. This suggests that the added autonomy and environmental access provided by agents is a key factor in achieving higher reproduction success rates, rather than model capability alone.

As ongoing work, this study naturally has limitations, and we have planned several next steps to address them. First, we consider expanding the benchmark to include additional studies, a broader range of failure modes, and more diverse analytical patterns common in computational social science. Second, we plan to evaluate additional models and agent architectures to better understand how general these findings are across different AI systems. Third, in the current agent-based setup, the ground-truth outputs are mounted within the same container used for repair. We verified through manual log inspection that agents did not access these outputs prematurely, but this separation is not enforced architecturally. Future work should isolate the validation environment from the repair environment to provide stronger guarantees against data leakage. Fourth, we will focus on systematically analyzing failure cases, particularly instances where automated workflows produce executable but incorrect results, to better characterize the limits of our approach.

\section{Data Availability}

All data and code used in this study are made publicly available. The repository for both workflows can be found at \href{https://github.com/Mehtab07/Automating-Computational-Reproducibility}{\nolinkurl{https://github.com/Mehtab07/Automating-Computational-Reproducibility}}

\vspace{0.5\baselineskip}
\noindent This work was supported by the German Research Foundation (DFG), project nos. 551687338 and 460234259. Artificial intelligence tools were used to assist with language editing and clarity during the preparation of this manuscript.

\bibliographystyle{ACM-Reference-Format}
\bibliography{reference}

\newpage
\appendix
\section{Code Snippet Example}
\label{app:sample}

This appendix provides an illustrative example of a corrupted code sample used in our experiments, showing the original script and a modified version containing an injected error.

\definecolor{newpurpleblue}{RGB}{138, 115, 226} 
\definecolor{skyblue}{RGB}{64, 156, 255}
\definecolor{purpleblue}{RGB}{120, 80, 220}
\definecolor{mintgreen}{RGB}{102, 255, 178}
\definecolor{sunsetorange}{RGB}{255, 140, 50}
\definecolor{lavender}{RGB}{180, 130, 255}

\begin{tcolorbox}[colback=blue!5!white,colframe=blue!50!black,title=Original Script,boxrule=0.5mm,enhanced]
\begin{lstlisting}[language=R,
                   basicstyle=\ttfamily\small,
                   keywordstyle=\ttfamily, 
                   commentstyle=\ttfamily\small,
                   stringstyle=\ttfamily\small,
                   morekeywords={},
                   breaklines=true,
                   breakindent=0pt
                  ]
linterp <- function(x, y) {
  a <- coef(lm(y ~ x))[2]
  x.min <- x[1]; x.max <- x[length(x)]
  y.out <- numeric(length(x))
    #...rest of computation...
  return(y.out)
}
\end{lstlisting}
\end{tcolorbox}

\begin{tcolorbox}[colback=red!5!white,colframe=red!50!black,title=Injected Error,boxrule=0.5mm,enhanced]
\begin{lstlisting}[language=R,
                   basicstyle=\ttfamily\small,
                   keywordstyle=\ttfamily, 
                   commentstyle=\ttfamily\small,
                   stringstyle=\ttfamily\small,
                   morekeywords={}
                   breaklines=true,
                   breakindent=0pt
                  ]
linterp <- function(x, y, x.out) {
  stop("Not implemented")
}
\end{lstlisting}
\end{tcolorbox}

\section{Templates for Prompt-based Repair}
\label{app:prompts}

Below are the templates for the prompts used for the three different context levels in the prompt-based repair workflow. These prompts were programmatically filled and instantiated with the respective synthetic test cases.

\begin{tcolorbox}[colback=skyblue!10!white, colframe=purpleblue!60!black, 
                  title=Minimal Context Prompt, 
                  boxrule=0.5mm, arc=3mm,enhanced, breakable]
\begin{lstlisting}[language=Python,
                   basicstyle=\ttfamily\small,
                   keywordstyle=\ttfamily, 
                   commentstyle=\ttfamily\small,
                   stringstyle=\ttfamily\small,
                   morekeywords={},
                   breaklines=true,
                   breakindent=0pt
                  ]
You are an AI assistant helping to fix R script that has failed.

--- Error Log ---
{log}

--- R Script ({script_name}) ---
{script_code}

Fix the script so it runs successfully. Add#comments beside, to explain your changes.
VERY IMPORTANT: Return only the corrected  R code. Do not include any explanatory text, markdown formatting, or any other tags like <think> or ```r.
\end{lstlisting}
\end{tcolorbox}
\vspace{1cm}
\begin{tcolorbox}[colback=skyblue!10!white, colframe=purpleblue!60!black, 
                  title=Medium Context Prompt, 
                  boxrule=0.5mm, arc=3mm, enhanced, breakable]
\begin{lstlisting}[language=Python,
                   basicstyle=\ttfamily\small,
                   keywordstyle=\ttfamily, 
                   commentstyle=\ttfamily\small,
                   stringstyle=\ttfamily\small,
                   morekeywords={},
                   breakatwhitespace=false,
                   breaklines=true,
                   breakindent=0pt
                  ]
You are an AI assistant helping to fix an  R script using paper context and error log.Below is an R script that is either incomplete or failing due to errors. Error logs (below) are provided to help you understand the issues. Use the paper's context (below) to understand what the script is trying to do and fix any issues, including missing code, incorrect file paths, masked out or NotImplemented functions, or undefined functions or dependencies issues.

--- Paper (Markdown) ---
{paper}

--- Error Log ---
{log}

--- R Script ({script_name}) ---
{script_code}

Fix the script so it runs successfully. Add#comments beside, to explain your changes.
VERY IMPORTANT: Return only the corrected  R code. Do not include any explanatory text, markdown formatting, or any other tags like <think> or ```r.
\end{lstlisting}
\end{tcolorbox}

\begin{tcolorbox}[colback=skyblue!10!white, colframe=purpleblue!60!black, 
                  title=Full-context Prompt, 
                  boxrule=0.5mm, arc=3mm, enhanced, breakable]
\begin{lstlisting}[language=Python,
                   basicstyle=\ttfamily\small,
                   keywordstyle=\ttfamily,
                   commentstyle=\ttfamily\small,
                   stringstyle=\ttfamily\small,
                   morekeywords={},
                   breaklines=true,
                   breakatwhitespace=false,
                   breakindent=0pt
                  ]
You are an AI assistant fixing an R script using full-context. You are an AI R programmer helping with computational reproducibility.

You are given:
1. The R script (below), which contains errors including missing code, incorrect file paths, masked out or NotImplemented functions, or undefined functions or dependencies issues.
2. Several other R scripts, which may help with context.
3. A markdown version of the research paper to understand what the function should do.
4. The error log from the last time the script was run.

Your task: Read the log file, inspect the script to modify and understand the error, and fix the error in the file listed below using the provided context. The function  to implement is in the file: {script_name}

--- Paper (Markdown) ---
{paper}

--- Other R Scripts (Context) ---
{context}

--- Last Run Error Log ---
{log}

--- Script to Modify ({script_name}) ---
{script_code}

--- Task ---
If there are NotImplemented functions, implement them **in-place** within the script above. Use the paper, other scripts,and log to help.

Fix the script so it runs successfully, using context from the paper and other scripts. Add #comments beside, to explain your changes. 

VERY IMPORTANT: Return only the corrected  R code. Do not include any explanatory text, markdown formatting, or any other tags like <think> or ```r.
\end{lstlisting}
\end{tcolorbox}

\section{Prompt for Agent-based workflow}
\label{app:prompt}

\begin{tcolorbox}[colback=skyblue!10!white, colframe=purpleblue, 
                  boxrule=0.5mm, arc=3mm,enhanced, breakable]
\begin{lstlisting}[language=Python,
                   basicstyle=\ttfamily\small,
                   keywordstyle=\ttfamily,
                   commentstyle=\ttfamily\small,
                   stringstyle=\ttfamily\small,
                   morekeywords={},
                   breaklines=true,
                   breakindent=0pt
                  ]
You are an autonomous research assistant responsible for repairing and validating R scripts for computational reproducibility.

**IMPORTANT: You must follow all the steps in the workflow. Do not stop until you have created the `status.txt` file.**

Environment Details:
  - Your working directory `/workspace` contains:
  - One or more R scripts to be run.
  - A research paper describing the expected outputs.
  - Associated data files and dependency files.
  - R script that currently fail when executed.
  - A directory named `base_results` is mounted at the root of the container. **Its absolute path is `/base_results`**. Do not look for it in the current directory. It contains the original expected output from the paper. Do *not* open or reference `/base_results` until after the repaired script runs successfully.

Your Task Workflow:
1.  First, create a plan. Write down the steps you will take to solve the problem.
2.  Identify the R script that fails to execute.
3.  Run the script and observe the error. Log all error output to console.
4.  Locate the source of the failure (e.g.,missing library, path issue, syntax error, incorrect variable name).
5.  Apply the minimal necessary fix directly in the original script. Examples:
   - If a library is missing, add `install.packages("library_name")` at the top of the script.
   - If a file path is incorrect, correct only the necessary part do not restructure the project.
   - If a there is a missing code of not implemented function, write that part instead of just commenting out those lines.
6.  After making the fix, run the script again to confirm whether it executes fully from start to finish.
7.  If new errors appear, iteratively fix them using the same minimal modification principle but do not edit anything other than the reason of actual error.

Reproducibility Check (Performed Only  After Successful Execution):
1.  Compare the output generated by the fixed script with the reference output in `/base_results`.
2.  Determine reproducibility status:
   - "Reproduced": Outputs match exactly.
   - "Not Reproduced": Outputs are missing, incomplete, or different.

Final Deliverables:
- Print a clear summary of the changes you made and why you made them.
- Create a new file named `status.txt` containing only one of the following words:
  - `Reproduced`
  - `Not Reproduced`
- The `status.txt` file must not contain any other text, characters, or emojis.
- Finally, print the message "Workflow complete."

Guidelines:
- Prefer minimal, targeted changes.
- Do not rewrite the entire script.
- Do not hallucinate results or outputs; your job is to make the code run, not to invent findings.
- Log all edits and actions clearly in the workspace.

You may now begin.
\end{lstlisting}
\end{tcolorbox}

\end{document}